\documentclass[amsmath,amssymb,aps,twocolumn,showpacs,prl,floatfix]{revtex4}

\usepackage{mathrsfs}
\usepackage{amsmath}
\usepackage{amssymb}
\usepackage{color}
\usepackage{graphicx}	
\usepackage{bm}
\usepackage{ulem} 

\newcommand{\be}{\begin{equation}}
\newcommand{\ee}{\end{equation}}
\newcommand{\bef}{\begin{figure}}
\newcommand{\eef}{\end{figure}}
\newcommand{\bea}{\begin{eqnarray}}
\newcommand{\eea}{\end{eqnarray}}

\begin{document}
\title{Numerical analysis of Pickering emulsion stability: insights from ABMD simulations}
\author{Fran\c cois Sicard}
\thanks{Corresponding author: \texttt{francois.sicard@free.fr}.}
\author{Alberto Striolo}

\affiliation{Department of Chemical Engineering, University College London, Torrington Place, London WC1E 7JE, United Kingdom, EU}
\date{\today}
\begin{abstract}
The issue of the stability of Pickering emulsions is tackled at a mesoscopic level 
using dissipative particle dynamics simulations within the Adiabatic Biased Molecular 
Dynamics framework. We consider the early stage of the coalescence process between two spherical water droplets 
in decane solvent. The droplets are stabilized by Janus nanoparticles of different shapes (spherical and ellipsoidal) 
with different three-phase contact angles. Given a sufficiently dense layer of particles on the droplets, 
we show that the stabilization mechanism strongly depends on the collision speed. This is consistent 
with a coalescence mechanism governed by the rheology of the interfacial region.
When the system is forced to coalesce \textit{sufficiently slowly}, we investigate at a mesoscopic level how 
the ability of the nanoparticles to stabilize Pickering emulsions is  discriminated by nanoparticle mobility 
and the associated caging effect. These properties are both related to the interparticle interaction and the hydrodynamic 
resistance in the liquid film between the approaching interfaces.
\end{abstract}


\maketitle

\section{Introduction}
Pickering emulsions~\cite{Pickering1907}, \textit{i.e.}, particle-stabilized emulsions, have been used in various 
applications including biofuel processing~\cite{Drexler2012}, encapsulation for drug delivery~\cite{Angelova2011},  
and food preservation~\cite{Shchukina2012,Puglia2012}.
When nanoparticles (NPs) are used as emulsifiers, their characteristics might have critical effects 
on properties such as interfacial tension~\cite{Miller2006}, droplet size, 
and emulsion stability~\cite{Wang2011,Zahn1997,Cheung2010,Weeks2002,Sonnenburg1991,Peng2009,Cheng2012}.
Although the factors affecting Pickering emulsion stability are still a key topic of research, 
the density of NPs on the droplet surface is known to be among the most important parameters~\cite{LuuJPCB2013}.
Indeed, the interfacial tension of the droplet depends strongly on the NP coverage~\cite{LuuJPCB2013}, 
which also depends on the NP affinity to the interface, \textit{i.e.}, the NP desorption energy. 
When the NP density on the droplet is large enough, the three-phase contact angle discriminates 
between NPs that are effective at stabilizing emulsion and those that are not~\cite{FanSM2012, FanPRE2012}. 
It is known that particles, once adsorbed, are rather difficult to displace from the interface. 
It is this property that makes particles such a good colloidal stabilizer for emulsions and foams. 
Besides the structure of NP monolayers, the NP diffusion is also believed to affect the system stability. 
Janus nanoparticles are believed to be more effective than homogeneous NPs in the stabilization of emulsion~\cite{FanSM2012}. 
Indeed, Janus NPs typically have high adsorption energies, and therefore are expected to pack densely at an interface.\\

In our present work, we study the early stage of the coalescence process between two spherical water droplets 
in decane solvent. The droplets are stabilized with various Janus NP types. We implement 
Adiabatic Biased Molecular Dynamics (ABMD) simulations~\cite{Paci1999,Camilloni2011, Marchi1999}, 
where the system is forced to coalesce \textit{adiabatically}. We show that, for a sufficiently dense 
layer of particles adsorbed on water droplets, the stabilization mechanism can strongly depend on 
the collision speed achieved during the merging process.
For collision velocities sufficiently slow compared to the rate of drainage of the thin liquid film 
trapped between the coalescing emulsion droplets, the ability of the NPs to stabilize Pickering emulsions 
is dictated by NP mobility and the associated caging effect.
This coalescence process is compared with simulations at sufficiently high collision speed, 
which suggest the droplet faculty to undergo large elastic deformation before coalescing is the stabilization 
mechanism. 

\section{Methods and algorithms}
\subsection{MD simulation}
The Dissipative Particle Dynamics (DPD) simulation method~\cite{Groot1997} was implemented within the simulation package 
LAMMPS~\cite{Plimpton1995}. The procedure and the parametrisation details are fully described in prior 
work~\cite{LuuLangmuir2013, LuuJPCB2013}. The system simulated here is composed of water, oil (decane), 
and NPs. One "water bead" (w) represents 5 water molecules. One decane molecule is modeled as two "oil beads" (o) 
connected by one harmonic spring of length $0.72$ $R_c$ and spring constant $350$ $k_BT/R_c$~\cite{Groot2001},
where $R_c$ is the DPD cutoff distance. The size of the simulation box 
is $L_x \times L_y \times L_z \equiv 72 \times 72 \times 140$ $R_c^3$, where $L_i$ is the box length 
along the $i^{th}$ direction. Periodic boundary conditions are applied in all three directions.
The NPs are modelled as hollow rigid ellipsoids and contain polar (p) and nonpolar (ap) DPD beads 
on their surface. One DPD bead was placed at the NP center for convenience, as described 
elsewhere~\cite{LuuLangmuir2013, LuuJPCB2013}. Hollow models have been used in the literature to simulate NPs, 
and hollow NPs can also be synthesized experimentally~\cite{Calvaresi2009}. 
We considered spherical and ellipsoidal NP with shape defined by the equation $x^2/b^2 + y^2/b^2 +z^2/c^2=1$, 
where $x$, $y$, and $z$ are Cartesian coordinates and $c$ and $b$ are the semi-axes of the ellipsoid. 
When $b=c$, spherical NPs are obtained. When $c<b$, the ellipsoid is oblate; when $b<c$, it is prolate. 
All NPs simulated had the same volume, $4/3 \pi a_0^3$, where $a_0$ is the radius of the equivalent sphere. 
We imposed $a_0 = 2R_c \approx 1.5$ nm.
All types of beads in our simulations have reduced mass of $1$. The total number of beads on one NP surface 
changes with the aspect ratio $c/b$. This allows us to maintain the surface bead density sufficiently high 
to prevent other DPD beads (either decane or water) from penetrating the NPs (which would be unphysical), as 
it has already been explained elsewhere~\cite{LuuJPCB2013}. To differenciate every NPs, we report 
the nonpolar fraction $N$ of the NP surface beads. For example, $N=70$ indicates that $70\%$ of the beads 
on the NP surface are nonpolar.
The interaction parameters shown in Table \ref{Tab-interaction} are used herein. These parameters were 
adjusted to reproduce selected atomistic simulation results, as explained in prior work~\cite{LuuLangmuir2013}. By tuning 
the interaction parameters between polar or nonpolar NP beads and the water and decane beads present in 
our system, it is possible to quantify the effect of surface chemistry on the structure and dynamics of NPs 
at water-oil interfaces.
\begin{table}[b]
\begin{center}
\begin{tabular*}{0.45\textwidth}{@{\extracolsep{\fill}}ccccc}
  \hline
  {} & $w$ & $o$ & $ap$ & $p$ \\
  \hline
  $w$ & $131.5$ & $198.5$ & $178.5$ & $110$  \\
  $o$ & {} & $131.5$ & $161.5$ & $218.5$ \\
  $ap$ & {} & {} & $450$ & $670$ \\
  $p$ & {} & {} & {} & $450$ \\
  \hline
\end{tabular*}
\caption{DPD interaction parameters expressed in $k_BT/R_c$ units. Symbols $w$, $o$, $ap$, and $p$ stand for 
water beads, oil beads, NP nonpolar beads, and NP polar beads, respectively.}
\label{Tab-interaction}
\end{center}
\end{table}
All simulations were carried out in the NVE ensemble~\cite{LuuLangmuir2013}. The scaled temperature was $1$, 
equivalent to $298.73$ K. 
The DPD time scale can be gauged by matching the self-diffusion of water. As demonstrated by Groot and Rabone \cite{Groot2001}, 
the time constant of the simulation can be calculated as $\tau = \frac{N_m D_{\textrm{sim}} R_c^2}{D_{\textrm{water}}}$, where $\tau$ is 
the DPD time constant, $D_{\textrm{sim}}$ is the simulated water self-diffusion coefficient, and $D_{\textrm{water}}$ is the experimental 
water self-diffusion coefficient. When $a_{w-w} = 131.5$ $k_B T/R_c$ (cf. Tab. \ref{Tab-interaction}), 
we obtained $D_{\textrm{sim}} = 0.0063$ $R_c^2/\tau$. For $D_{\textrm{water}} = 2.43 \times 10^{-3}$ $cm^2/s$ \cite{Partington1952}, 
we finally obtain $\tau = 7.6$ ps.
In the simulations discussed herein, the size of the droplet was fixed. At the beginning of each simulation, 
the solvent (oil) beads were distributed within the simulation box forming a cubic lattice. 
Two water droplets of radius $\approx 20$ $R_c$ were generated by replacing the oil beads with water beads 
within the volume of two spherical surfaces separated with a distance of $50$ $R_c$. 

A number of spherical or ellipsoidal NPs were placed randomly at the water-decane interface 
with their polar (nonpolar) part in the water (oil) phase to reach the desired water-decane interfacial area per NP. 
Following previous work~\cite{LuuJPCB2014}. the NPs considered in this study are oblate with an aspect ration $c/b = 1$ when 
spherical and $c/b=2$ when ellipsoidal. Note that nonspherical NPs have been considered as it has been reported 
they can be more efficient in stabilizing emulsions than spherical NPs.
Indeed, it has been reported that the minimum amount of particle needed to stabilize an emulsion decreases as the particle aspect ratio increases~\cite{Mejia2012, Dendukuri2006}. 
The initial configuration obtained was simulated for $10^6$ timesteps in order to relax the density of the system and the contact angle of 
the nanoparticles on the droplet. The system pressure and the three-phase contact angles did not change notably 
after 5000 simulation steps.

\begin{table*}
\resizebox{\textwidth}{!}{%
\begin{tabular}{|c|c|c|c|c|c|}
   \cline{3-6}
     \multicolumn{2}{c|}{} & $\textbf{144}$ \textbf{NPs} & $\textbf{160}$ \textbf{NPs} & $\textbf{165}$ \textbf{NPs} & $\textbf{170}$ \textbf{NPs}\\
    \hline
				& $\mathcal{A}_{NP}$ $(R_c^2)$ & $30.84 \pm 0.06 $ & $27.61 \pm 0.06 $ & $26.75 \pm 0.08$ & $25.96 \pm 0.13$ \\
    				& $\theta_C$  & $91.4^\circ \pm 2.1^\circ$ & $91.3^\circ \pm 2.1^\circ$ & $91.6^\circ \pm 2.2^\circ$ & $91.6^\circ \pm 2.3^\circ$ \\ 
\textbf{S-55JP}	& $R_G$ $(R_c)$ & $13.951 \pm 0.003$ & $13.824 \pm 0.003$ & $13.777 \pm 0.003$ & $15.562 \pm 0.004$ \\ 
				& $As$ $(R_c)$ & $1.54 \pm 0.28$ & $1.68 \pm 0.24$ & $1.70 \pm 0.26$ & $2.81 \pm 0.16$\\
				& $An$ & $(2.4 \pm 1.6)\times10^{-4}$ & $(2.13 \pm 1.29)\times10^{-4}$ & $(7.18 \pm 3.64)\times10^{-4}$ & $(12.6 \pm 2.8)\times10^{-4}$\\
    \hline
				& $\mathcal{A}_{NP}$ $(R_c^2)$ & $31.11 \pm 0.07 $ & $27.82 \pm 0.06 $ & $26.89 \pm 0.07$ & $26.10 \pm 0.10$ \\
				& $\theta_C$  & $95.2^\circ \pm 2.4^\circ$ & $95.4^\circ \pm 2.3^\circ$ & $95.3^\circ \pm 2.6^\circ$ & $95.3^\circ \pm 2.5^\circ$ \\
\textbf{S-60JP}	& $R_G$ $(R_c)$ & $13.937 \pm 0.003$ & $13.810 \pm 0.003$ & $13.761 \pm 0.003$ & $13.728 \pm 0.003$ \\
				& $As$ $(R_c)$ & $1.53 \pm 0.28$ & $1.54 \pm 0.29$ & $1.65 \pm 0.24$ & $2.35 \pm 0.18$ \\
				& $An$ &  $(2.42 \pm 1.49)\times10^{-4}$ & $(2.71 \pm 1.75)\times10^{-4}$ & $(2.93 \pm 1.34)\times10^{-4}$ & $(12.7 \pm 3.2)\times10^{-4}$ \\
     \hline 
				& $\mathcal{A}_{NP}$ $(R_c^2)$ & $31.44 \pm 0.08 $ & $28.14 \pm 0.05 $ & $27.21 \pm 0.05$ & $26.29 \pm 0.06$ \\       
				& $\theta_C$  & $100.6^\circ \pm 2.8^\circ$ & $100.5^\circ \pm 2.8^\circ$ & $100.6^\circ \pm 3.2^\circ$ & $101.0^\circ \pm 3.2^\circ$ \\
\textbf{S-70JP}	& $R_G$ $(R_c)$ & $13.921 \pm 0.003$ & $13.806 \pm 0.003$ & $13.771 \pm 0.003$ & $13.703 \pm 0.004$ \\
				& $As$ $(R_c)$ & $1.53 \pm 0.25$ & $1.54 \pm 0.25$ & $1.90 \pm 0.25$ & $1.85 \pm 0.25$ \\
				& $An$ & $(2.17 \pm 1.06)\times10^{-4}$ & $(2.70 \pm 1.42)\times10^{-4}$ & $(5.42 \pm 2.46)\times10^{-4}$ & $(6.18 \pm 3.28)\times10^{-4}$ \\
     \hline \hline  
				& $\mathcal{A}_{NP}$ $(R_c^2)$ & $31.07 \pm 0.11 $ & $27.67 \pm 0.09 $ & $26.75 \pm 0.09$ & $25.93 \pm 0.09$ \\       
				& $\theta_a$  & $61.2^\circ \pm 7.6^\circ$ & $56.7^\circ \pm 10.1^\circ$ & $52.8^\circ \pm 10.8^\circ$ & $51.6^\circ \pm 11.3^\circ$ \\
\textbf{E-70JP}	& $R_G$ $(R_c)$ & $14.035 \pm 0.003$ & $13.805 \pm 0.004$ & $15.812 \pm 0.004$ & $13.649 \pm 0.004$ \\
				& $As$ $(R_c)$ & $1.61 \pm 0.30$ & $1.97 \pm 0.33$ & $1.97 \pm 0.33$ & $1.69 \pm 0.27$ \\
				& $An$ & $(3.02 \pm 1.92)\times10^{-4}$ & $(5.92 \pm 3.27)\times10^{-4}$ & $(3.77 \pm 2.06)\times10^{-4}$ & $(3.86 \pm 2.19)\times10^{-4}$ \\
     \hline         
\end{tabular}
}%
\caption{Interfacial area per NP ($\mathcal{A}_{NP}$), contact angles ($\theta_C$), average orientation angles ($\theta_a$),
 radius of gyration ($R_G$), asphericity ($As$), and relative shape anisotropy ($An$), obtained from 
simulating the various NPs at the decane-water interface. In all cases, the water droplets were simulated in decane solvent.
The statistical errors are estimated as one standard deviation from the average contact angles obtained 
for equilibrated trajectories.}
\label{Tab-characteristic}
\end{table*}

\subsection{Droplet characteristics}
It is well established that the stability of Pickering emulsions is chiefly determined by the size, topology, 
and concentration of particles at the interface, as well as their wetting properties~\cite{FanSM2012, LuuJPCB2013,LuuJPCB2014,Qi2014,French2015}. 
In the following we consider a system made by two identical spherical water droplets immersed in oil, 
and stabilized by a sufficiently dense layer of NPs. 
By \textit{sufficiently dense} layer of NPs, we mean the NP concentration must be dense enough that 
the liquid film drainage time, \textit{i.e.}, the time interval from  droplet contact to fusion, is significant.
This can be tackled in a first instance considering \textit{steered} simulations with a range of pulling 
rates representative of shear rates that can be reached in a conventional mechanical 
emulsification~\cite{Mason1996} or ultrasonication processes~\cite{Delmas2011}.
In the following, we consider a limited range of NP concentrations and types (\textit{i.e.} contact angles) 
as well as different shape (spherical and ellipsoidal), as initial conditions.
Table~\ref{Tab-characteristic} summarizes the thermodynamic (three-phase contact angle, NP orientation angle)
and the geometrical properties~\cite{Vymetal2011} (radius of gyration, asphericity, and relative shape anisotropy of the droplet) 
for different NP types and concentrations that we consider throughout this study.

\textbf{Three-phase contact angle}. To estimate the three phase contact angle on the droplets we calculate 
the fraction of the spherical NP surface area that is wetted by water, and we obtained $\theta_C$ as
\begin{equation}
\theta_C = 180 - \arccos\Big(1-\frac{2 A_w}{4\pi R^2}\Big) \, ,
\end{equation}
where $A_w$ is the area of the NP surface that is wetted by water and $R$ is the radius of the NP. The ratio $A_w/4\pi R^2$ 
is obtained by dividing the number of NP surface beads (ap or p), which are wetted by water, by the total number of beads 
on the NP surface ($192$ for spherical NP). One surface bead is wet by water if a water bead is the solvent bead nearest to it. 
One standard deviation from the average is used to estimate the statistical uncertainty. 

\textbf{Orientation angle}. To characterize the ellipsoidal NPs structure on a droplet, we focus on the orientation 
of their longest axes (the $c$ axes) with respect to the radial directions from the center of the droplet~\cite{LuuJPCB2014}. 
The orientation angle $\theta_a$ is defined as
\begin{equation}
\theta_a = \arccos(\textbf{u}_i \cdot \textbf{R}_i) \, ,
\end{equation}
where $u_i$ is the unit vector along the NP $c$ axis and $R_i$ is the unit vector representing the radial direction 
from the center of the droplet. When $\theta_a = 0$ $(\pi/2)$, the correspondent NP is parallel (perpendicular) to the droplet 
radial direction, and therefore perpendicular (parallel) to the local liquid-liquid interface. Note that we define the angle 
$\theta_a$ so that it is never larger than $\pi/2$. One standard deviation from the average is used to estimate the statistical uncertainty. 
\begin{table*}[t]
\begin{center}
\begin{tabular*}{0.9\textwidth}{@{\extracolsep{\fill}}ccccc}
\hline
{} & $144$ NPs & $160$ NPs & $165$ NPs & $170$ NPs \\
  \hline
  S-55JP & $18.8 \pm 0.2$ & $18.8 \pm 0.3$ & $18.7 \pm 0.4$ & $18.7 \pm 0.6$  \\
  S-60JP & $18.9 \pm 0.3$ & $18.8 \pm 0.3$ & $18.8 \pm 0.3$ & $18.8 \pm 0.5$ \\
  S-70JP & $19.0 \pm 0.3$ & $18.9 \pm 0.2$ & $18.9 \pm 0.3$ & $18.9 \pm 0.3$ \\
  \hline \hline
  E-70JP & $18.9 \pm 0.4$ & $18.8 \pm 0.4$ & $18.7 \pm 0.4$ & $18.7 \pm 0.4$ \\
  \hline
\end{tabular*}
\caption{Water droplet radius, $R_{drop}$, expressed in $R_C$ units, for various NP types and concentrations. 
The statistical errors are estimated as one standard deviation from the average obtained 
for equilibrated trajectories. $R_{drop}$ is the observable that allows us to fix the target value $S_{\textrm{target}}$ for the coordinate 
used in ABMD simulations.}
\label{Tab-radius}
\end{center}
\end{table*}

\textbf{Radius of gyration, asphericity and relative shape anisotropy}. The description of the geometrical 
properties of complex systems by generalized parameters such as the radius of gyration or principal components of the gyration 
tensor has a long history in macromolecular chemistry and biophysics~\cite{Vymetal2011,Solc1971}.
Indeed, such  
descriptors allow an evaluation of the overall shape of a system 
(spherical, oblate, or prolate) and reveal its symmetry. Considering, e.g., the following definition for the gyration tensor, 
\begin{equation}
\mathcal{T}_{GYR} = \frac{1}{N}
\begin{bmatrix}
\sum x_i^2 & \sum x_i y_i & \sum x_i z_i \\
\sum x_i y_i & \sum y_i^2 & \sum y_i z_i \\
\sum x_i z_i & \sum y_i z_i & \sum z_i^2
\end{bmatrix} \, ,
\end{equation}
where the summation is performed over $N$ atoms and the coordinates $x$, $y$, and $z$ are related to the 
geometrical center of the atoms, one can define a reference frame where $\mathcal{T}_{GYR}$ can be diagonalized:
\begin{equation}
\mathcal{T}_{GYR}^{diag} =
\begin{bmatrix}
S_1 & 0 & 0\\
0 & S_2 & 0 \\
0 & 0 & S_3
\end{bmatrix} \, .
\end{equation}
In this format we obey the convention of indexing the eigenvalues according to their magnitude. 
We thus define the radius of gyration $R_{GYR}^2 \equiv S_1^2 + S_2^2 + S_3^2$, 
the asphericity $A_s \equiv S_1 - \frac{1}{2}(S_2+S_3)$, which measures the deviation from the spherical 
symmetry, and the relative shape anisotropy $A_n^2 \equiv 1-3\frac{S_1S_2+S_1S_3+S_2S_3}{(S_1+S_2+S_3)^2}$, 
which reflects both the symmetry and dimensionality of a system~\cite{Vymetal2011}.

As discussed in the Electronic Supporting Information (ESI), we consider NP types and concentrations such that they are 
strongly effective at preventing droplet coalescence, and such that the shape of the droplet be spherical 
before contacting the neighbour droplet. In the simulation discussed herein, we consider spherical water droplets 
in decane solvent stabilized with 160-165 NPs.
\begin{figure*}[b]
\includegraphics[width=1.0 \textwidth, angle=-0]{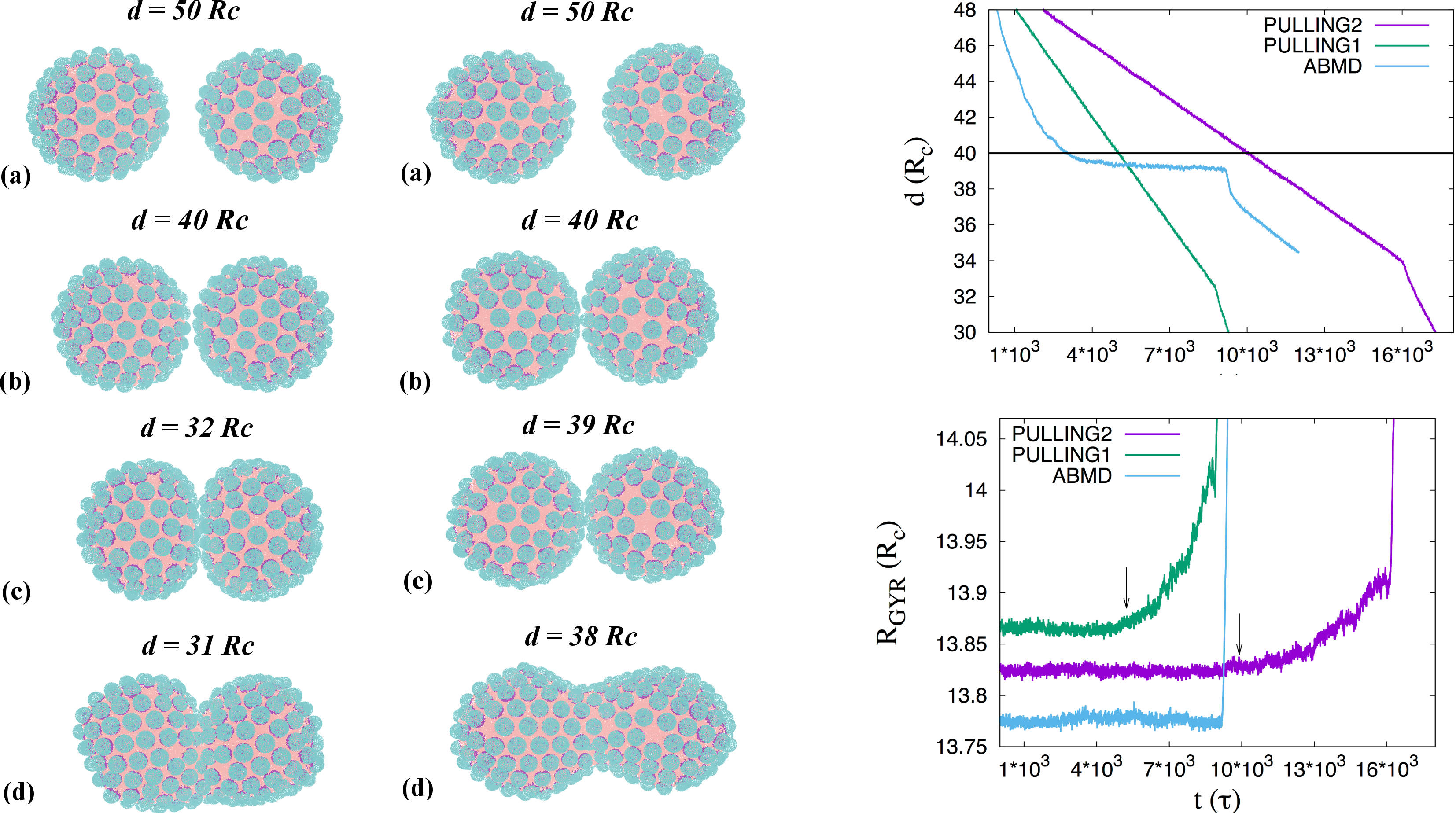}
 \caption{\textbf{Left panel}. Sequence of simulation snapshots representing typical coalescence processes of two 
 water droplets stabilized with 160 spherical nanoparticles of type 55JP. Green and purple spheres represent polar and apolar beads, respectively.
 The distance between the two droplets 
 is measured between their centers of mass and decreases from top to bottom. The left (right) 
  column represent the sequence of a pulling (ABMD) simulation where (a) is the initial distance, 
  (b) the first contact between the two droplets, (c) the coalescence threshold, and (d) the merging of 
  the two droplets. \textbf{Right panel}. Temporal evolution of the distance between the centers of mass of 
  the two droplets, $d$, (top) and the radius of gyration, $R_{GYR}$, of one incoming droplet (bottom) during 
  pulling and ABMD simulations. The temporal evolutions of $R_{GYR}$ are shifted arbitrarily along the 
  ordinate axis for clarity. The two arrows in the bottom panel indicate the departure from the initial plateau of $R_{GYR}$ 
  indicating the beginning of the shape alteration of the droplets. This corresponds to the distance $d = 40$ $R_C$ 
  (black plain line) at which the two coalescing droplets interact through NP-NP interaction. 
  We use the notation: Pulling1  $\big( v_P = 0.001$ $R_C$/time step, $k_P = 5000$ $k_BT/R_C^2 \big)$, 
  Pulling2 $\big( v_P = 0.0005$ $R_C$/time step, $k_P = 5000$ $k_BT/R_C^2 \big)$, 
  and ABMD $\big( \alpha = 1000$ $k_BT/R_C^2$, $S_{\textrm{target}}=39$ $R_C \big)$.}
\label{fig1}
\end{figure*}

\subsection{ABMD algorithm}
Adiabatic biased molecular dynamics~\cite{Marchi1999, Camilloni2011,Paci1999} (ABMD) is an algorithm developed to connect 
any two points in the conformational space of a given system. The method is based on the introduction 
of a biasing potential, which is a function of chosen coordinates of the system and which is zero 
when the system is moving toward the desired target point, while disfavoring motions in the opposite direction. 
This is similar to what happens in a \textit{ratchet-and-pawl} system, which undergoes random thermal fluctuations, 
while the pawl allows the ratchet to move only in one direction~\cite{Camilloni2011}. In ABMD the chosen coordinate 
plays the role of the ratchet and the biaising potential that of the pawl. 
In this respect, the biasing potential does not exert any work to direct the system toward the 
target conformation, as it would happen when pulling the system with a force; on the contrary the system makes moves 
toward the target conformation under the driving effect of the potential of the force-field alone. The biasing 
potential is implemented as
\begin{equation}
V\big(\rho(t)\big) = \left\{
    \begin{array}{ll}
        \frac{\alpha}{2} (\rho(t)-\rho_m(t))^2 & \rho(t) > \rho_m(t) \\
        0 & \rho(t) \leq \rho_m(t)
    \end{array}
\right.
\label{eq-ABMD}
\end{equation}
where 
\begin{equation}
\rho(t) = \big( S(t)-S_{\textrm{target}}\big)^2
\end{equation}
is the distance along the coordinate $S$ of the actual configuration of the system with respect to a target value 
$S_{\textrm{target}}$, $\alpha$ is a damping constant, and 
\begin{equation}
\rho_m(t) = \min_{0\leq \tau \leq t} \rho(\tau)
\end{equation}
is the minimum distance reached until time $t$. The algorithm is defined by the choice of the coordinate $S$ 
and of the damping constant $\alpha$.
The ABMD algorithm is routinely implemented by the biophysical community to explore free-energy paths 
associated to protein folding and unfolding~\cite{Paci1999,Provasi2010,Camilloni2011}. 
In such applications, the most probable sequence of events generated by 
the ABMD simulations is expected to coincide with the minimum free-energy path independently on the damping constant.
The outcome of a set of ABMD simulations is  a number of trajectories mapping the free-energy minimum of the system, 
 in which time and the statistical weight of the conformations along the trajectories are unphysically modulated. 
One of the unique features of this method is that the transformation connecting the two points of interest 
in the configurational space occurs following natural fluctuations. 
Additionally, the algorithm minimizes the sum of the potential and kinetic energy of the particles 
and of the bias potential. This provides means of controlling the quality and \textit{adiabaticity} of the simulation.

In this study, we use the efficiency of the ABMD algortihm within a different perspective, \textit{i.e.}, to  understand 
 the complex \textit{dynamic} processes of the early stage of droplet coalescence. The delicate issue is the choice of the 
damping constant $\alpha$ in Eq.~\ref{eq-ABMD} that drives the mechanism under study. According to the classical theory of fluid particle 
coalescence~\cite{Chan2011}, we aim at tackling at a mesoscopic scale the mechanism associated to the drainage of the liquid films 
of the continuous phase between two approaching droplets. These liquid films should drain to allow the 
dispersed fluids to make contact and fuse. The coalescence mechanism strongly depends on the velocity of the two approaching droplets, 
which can be responsible for large deformation of the droplet shape~\cite{FanSM2012}. In the following, we  consider ABMD parameters that 
allow us to study coalescence while keeping the shape of the droplets unaltered (\textit{i.e.}, the radius 
of gyration of the droplets). This means we shall use a value of $\alpha$ small enough so as to provide the correct 
sequence of events associated to coalescence, but large enough to make this happen in a computationally reasonable time. 
The second ABMD parameter of choice is the target value $S_{\textrm{target}}$, which is set on the distance between the two water droplet 
before coalescence occurs. $S_{\textrm{target}}$ is set equal to the radius of the isolated droplets, and determined by the positions 
of the NPs. These data are reported in Table.~\ref{Tab-radius}.
Biased simulations were performed using the version 2.1 of the plugin for free-energy calculation, 
named PLUMED~\cite{Bonomi2009,Tribello2014}.

\subsection{Pulling algorithm}
Pulling dynamics, also named steered molecular dynamics (SMD) is an algorithm which takes inspiration from 
single-molecule pulling experiments~\cite{Grubmuller1996} and forces a system to evolve away from its 
initial equilibrium condition. This is intended to accelerate transitions between different energy minima 
and to estimate the free-energy difference between two (or more) states via the Jarzynski equality~\cite{Park2003}. 
Pulling simulations have become very popular and are extensively applied in studying many biophysical 
processes~\cite{Isralewitz2001,Lu1998} and phenomena in chemical physics~\cite{FanSM2012}.
In SMD, the Hamiltonian of the system, $\mathcal{H}_0$, is modified into a new Hamiltonian, $\mathcal{H}$, 
which contains a term which depends on time via a harmonic spring potential centered on a point which 
moves linear with time, \textit{i.e.},
\begin{equation}
\mathcal{H} = \mathcal{H}_0 + \frac{k_P}{2}\Big(S(t)-S_0+v_Pt\Big)^2 \, ,
\end{equation}
where $S(t)$ is the chosen coordinate of the configuration of the system that is biased, $S_0 = S(t=0)$, 
$v_P$ the pulling velocity, and $k_P$ the spring constant. For an appropriate set of SMD parameters ($k_P$, $v_P$), 
the system follows closely the center of the moving harmonic spring. 

\section{Results and discussion}
In the following, we consider two types of numerical simulations, (1) \textit{Pulling simulations} 
where the two droplets are moved towards each other using one harmonic spring with a force constant 
of $5000$ $k_B T/R_c$ tethered to their center of mass, at constant relative velocity (pulling rate) 
$v_P = 0.001$ $R_c$ or $0.0005$ $R_c$ per time step, and (2) \textit{ABMD simulations} where 
we consider the distance between the center of mass of the two droplets as the biased coordinate $S$, 
a damping constant $\alpha = 1000$  $k_B T/R_c^2$ and a target value $S_{\textrm{target}} = 39$ $R_c$.

In Fig.\ref{fig1} (left panel) we show simulation snapshots obtained during the two coalescence simulations: 
pulling simulations (left column) and ABMD simulations (right column). In both cases we consider  water droplets 
stabilized by 160 NPs of type 55-JP in decane solvent. The interfacial area per NP equals $27.61$ $R_C^2$.
\begin{figure*}[b]
\includegraphics[width=1.0 \textwidth, angle=-0]{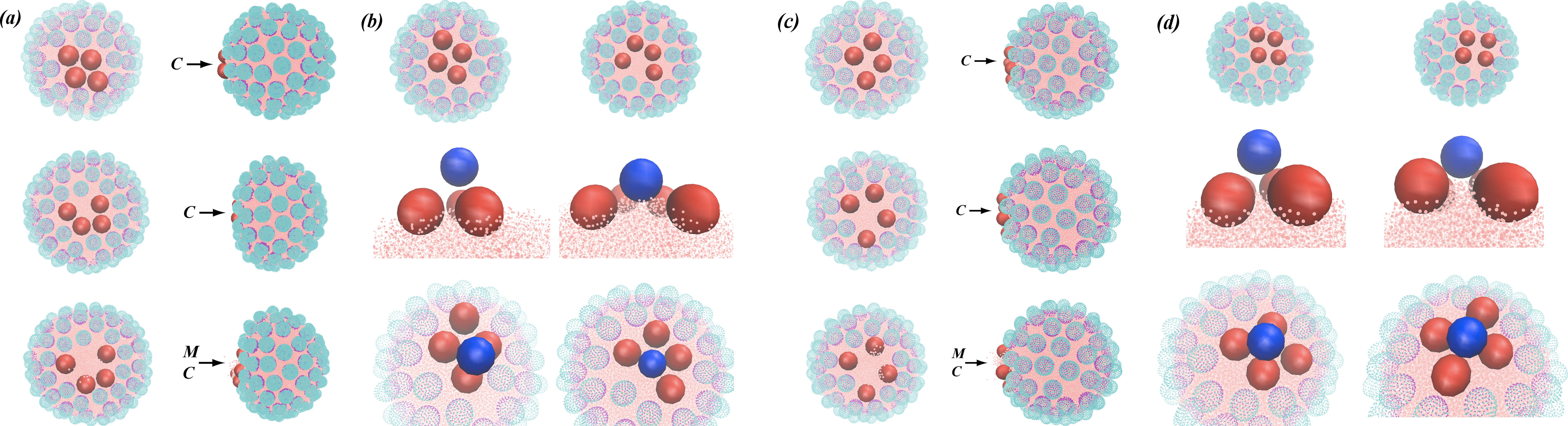}
 \caption{Sequence of simulation snapshots representing typical coalescence processes of two 
 water droplets stabilized with 160 spherical nanoparticles of type 55JP during Pulling (a,b) 
 and ABMD simualtions (c,d). Green and purple spheres represent polar and apolar beads, respectively.  
 $(M)$ and $(C)$ represent \textit{contact} and \textit{merging} 
 points between the incoming NP (blue sphere) and the set of NPs surrounding the neighbour droplet 
 that defines the interaction area (red spheres). The contact point (C) corresponds to the first NP-NP 
 interaction between the incoming NP and its neighbour droplet. The merging point (M) is where the water bridge 
 is formed between the two droplets. The interaction area is delimited by the set of NPs (red spheres) on the approaching droplet 
that defines the first neighbouring shell, \textit{i.e.}, the distance before merging between the center of the incoming (blue) NP 
and the centers of each red NP is less than $7$ $R_c$. In the configuration considered here, the merging point is located 
 in the interaction area, but this is not always the case (see Fig.~\ref{fig4}).}
\label{fig2}
\end{figure*}
Visual inspection of the simulation snapshots highlights the fundamental difference between the two processes. 
For the ABMD simulation, the shape of the water droplets remains unaltered, while in pulling simulation a clear alteration of 
the shape of the two coalescing droplets is observed as the distance between the center of mass of the droplets decreases. 
This behaviour is quantitatively shown in Fig.\ref{fig1} (right panel) where we show the temporal evolution of 
the distance between the centers of mass $d$  and the radius of gyration $R_G$ 
of the two coalescing droplets as a function of simulation time during ABMD and pulling simulations. 
Before the two droplets come into contact, the radius of gyration remains constant as the distance between the two droplets 
decreases. When the distance is such that the two droplets start interacting with each other via NP-NP interaction 
($d = 40$ $R_c$), $R_G$ increases
 during the pulling simulation, meaning the droplets deform as the distance imposed decreases. $R_G$ keeps 
 increasing until a minimal distance is reached ($d = 32$ $R_c$), when the two droplets coalesce. 
\begin{figure*}
\includegraphics[width=1.0 \textwidth, angle=-0]{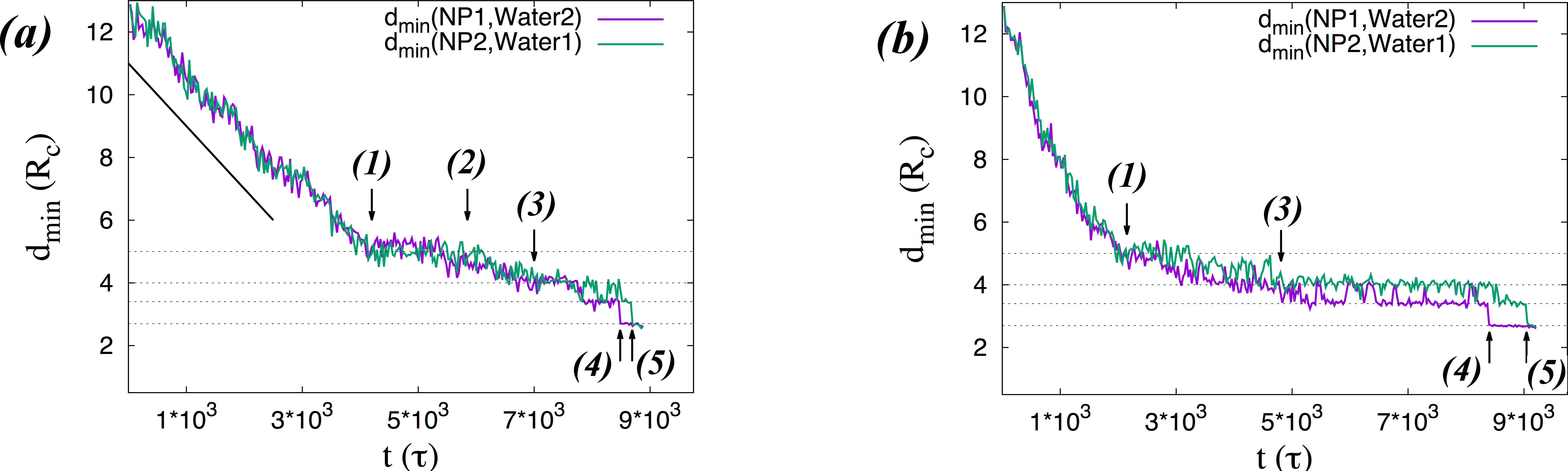}
 \caption{Temporal evolution of the minimal distance $d_{\min}$ between the center of the incoming droplet (NP1 or NP2) and the water 
 molecules from the other droplet (Water2 or Water1, respectively). Two water droplets stabilized with 160 spherical nanoparticles of type 55JP 
 in decane solvent are considered. During the Pulling simulation (panel a) the distance decreases 
 consistently with the pulling simulation rate (black plain line). The distance 
 $d_{\min}$ reaches a plateau (1) when the incoming NP interacts with a NP belonging to the second droplet. 
 The incoming NP then interacts with the solvent shell surrounding the droplet (2) and the water molecules (3). 
 The two droplets coalesce when a bridge is formed between two NPs from the two different droplets (4 and 5). 
 Considering the ABMD simulation (panel b), the distance $d_{\min}$ does not show the NP-NP interaction plateau (1). 
 In this system the incoming NP interacts with the solvent shell surrounding the droplet and the water molecules (3).
 As in the pulling simulation, the two droplets coalesce when a bridge is formed and facilitated by two NPs from 
 the two different droplets (4 and 5).
 }
\label{fig3}
\end{figure*}
If one considers the ABMD simulation, the distance between the two droplets reaches a plateau as soon as the two 
droplets come into contact, at around $d = 39$ $R_c$. This 
plateau is slightly lower than the value $d = 40$ $R_c$. This difference indeed can be explained by the fact that 
ABMD simulation  drives the system such that the NPs surrounding the droplet interact with the solvent shell that surrounds
the water molecules of the second droplet and thus avoid NP-NP interactions that are more energetically costly. 
The shape of the two droplets remains unaltered until 
the two droplets coalesce.

It is necessary to further quantify the difference between these two coalescence mechanisms at a mesoscopic scale.
In Fig.~\ref{fig2} we show detailed simulation snapshots obtained during both pulling and ABMD simulations. Considering first 
qualitatively the pulling simulation, the two droplets start interacting when one NP in the incoming droplet (blue sphere in right panel 
in Fig.~\ref{fig2}) comes into contact with one or more NPs from the opposite droplet (red spheres in Fig.~\ref{fig2}). 
This NP-NP interaction initiates the shape alteration of the interacting droplets. 
We identify an interaction area as the set of NPs on the neighbour droplet (red spheres in Fig.~\ref{fig2}) 
that defines the first neighbouring shell (\textit{i.e.}, distance between blue NP and red NPs less than $7$ $R_c$) 
around the incoming NP before merging. The $7$ $R_c$ threshold is set based on observations from prior work~\cite{LuuLangmuir2013}.
Then the incoming NP shifts towards the center of the interaction area delimited by the red  NPs 
in Fig.~\ref{fig2}. 
This is quantitatively described in Fig.~\ref{fig3} (left panel) where we follow the evolution of the minimal distance $d_{\min}$
between the center of one incoming droplet and the water molecules belonging to the other droplet. The first part of the evolution 
is representative of the pulling rate, until the distance reaches a plateau around $d_{\min}\approx 5$ $R_c$ corresponding 
to where the NP/NP interactions start. Then the incoming NP starts interacting with the solvent in the interacting area 
and keeps compressing the droplet. This evolution goes on until $d_{\min}$
reaches the minimal solvent shell, \textit{ie.}, $3.5 \leq d_{\min} \leq 4$ $R_c$, and the incoming NP starts coming into contact directly 
with the water molecules ($d_{\min} = 3$ $R_c$).
Merging of the two droplets occurs when two NPs (each belonging to different droplets) 
interact with water molecules belonging to the other droplet (cf. arrows (4) and (5) in Fig. \ref{fig2}). 
A water bridge is thus formed between the droplets, which allows the flow of water molecules.
\begin{figure*}
\begin{center}
\includegraphics[width=0.8 \textwidth, angle=-0]{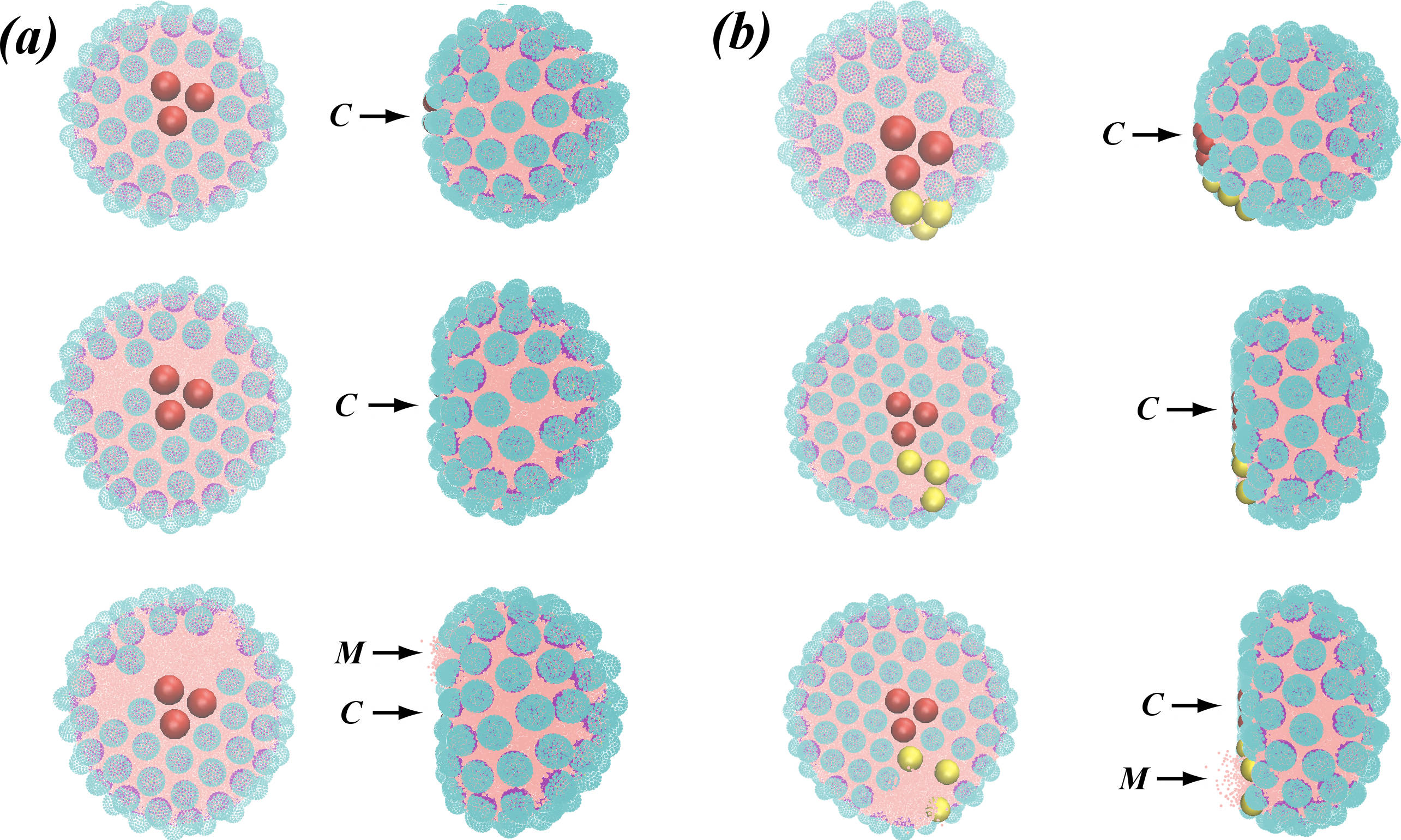}
 \caption{
 Sequence of simulation snapshots representing typical coalescence processes of two 
 water droplets stabilized with 160 spherical NPs (panel a) and 165 spherical NPs (panel b) of type 60JP 
 during pulling simulation. Green and purple spheres represent polar and apolar beads, respectively.  
 $(M)$ and $(C)$ represent  \textit{contact} and \textit{merging} 
 points, respectively, between one NP on the incoming droplet (blue sphere) and the interaction area delimited by the NPs highlighted in red
 on the second droplet (see main text and Fig.~\ref{fig2} for definitions).
 In panel (a), the droplet deforms generating a rather large interfacial area not covered by NPs. 
 This interfacial area favours the coalescence. In panel (b), the increase of the NP density 
 prevents the occurence of the large NP-free interfacial area. One NP on the incoming droplet finds itself in contact 
 with the water molecules from the other droplet (yellow spheres), favouring coalescence. 
 This NP, which effectively initiates coalescence, is different from the NP that induced the compression of the second droplet when the two droplets initially came into contact.
 }
\label{fig4}
\end{center}
\end{figure*}
\begin{figure*}
\begin{center}
\includegraphics[width=0.85 \textwidth, angle=-0]{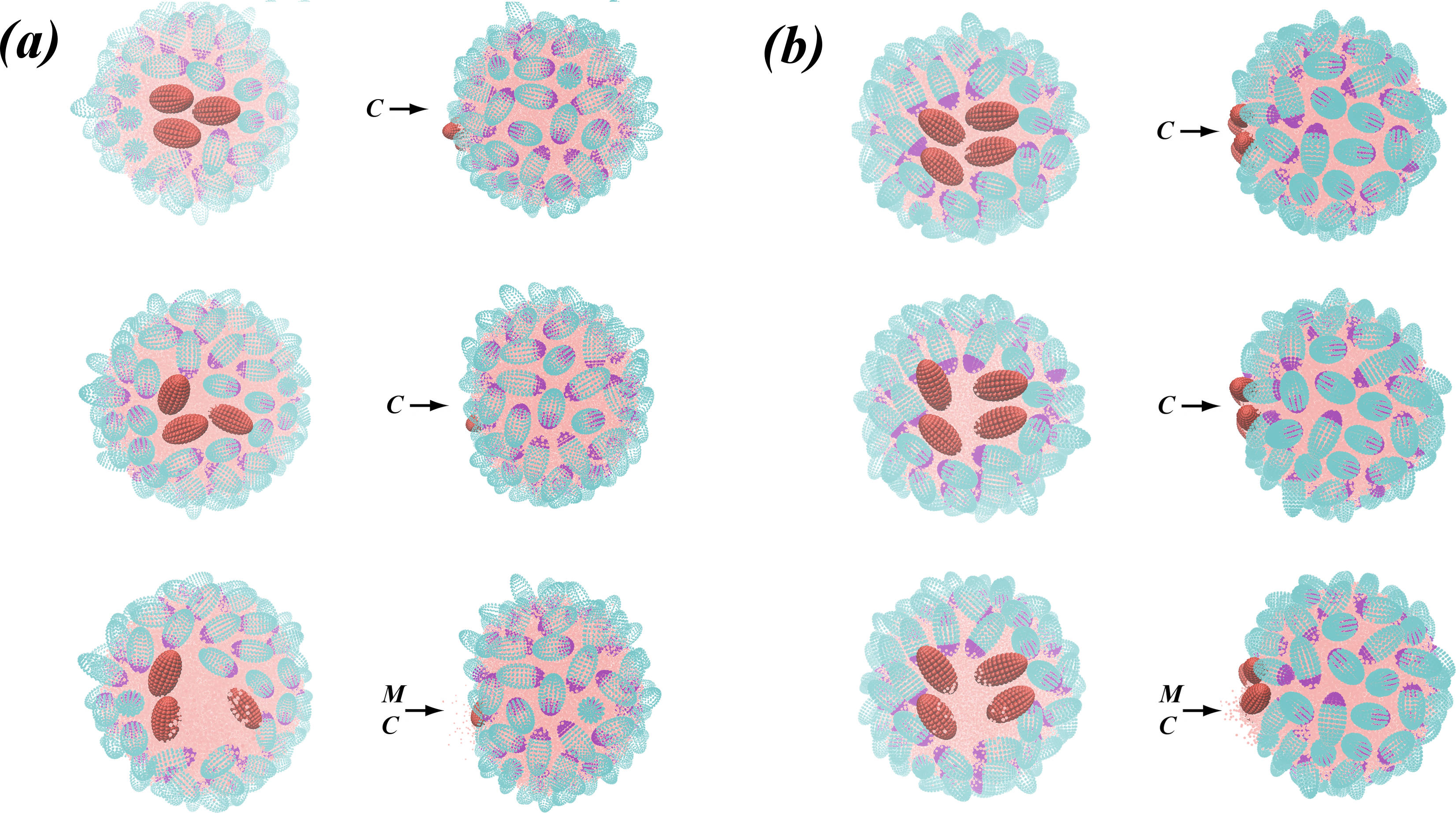}
 \caption{
 Sequence of simulation snapshots representing typical coalescence processes of two 
 water droplets stabilized with 160 ellipsoidal nanoparticles of type 70JP during pulling (panel a) 
 and ABMD simualtions (panel b). Green and purple spheres represent polar and apolar beads, respectively.
 $M$ and $C$ represent  \textit{contact} and \textit{merging} 
 points between one NP on the incoming droplet and a set of NPs on the second  droplet highlighted in red. These red spheres define 
 the interaction area (see main text and Fig.~\ref{fig2} for definitions).
 }
\label{fig5}
\end{center}
\end{figure*}

Let us now compare the behaviour just described with the one observed running the ABMD simulations. Considering the simulation snapshots 
in Fig.~\ref{fig2} and the quantitative analysis in Fig.~\ref{fig3}, we first see that the system does not spend 
a significant amount of time in the plateau at $d_{\min}\approx 5$ $R_c$. Analysis of the trajectories suggest that the NP 
in the incoming droplet avoids 
 frontal interactions with NPs belonging to the other droplet. 
Indeed, as prescribed by the ABMD algorithm, the system follows a minimal free-energy path such that the incoming droplet 
 mainly interacts with the solvent molecules located around the interaction area delimited by the red subset of 
 NPs in Fig.~\ref{fig2}.
The system  reaches progressively the plateau corresponding to the minimal solvent shell, \textit{i.e.}, $3.5 \leq d_{\min} \leq 4$ $R_c$,
and the incoming NP eventually comes into contact with the water molecules ($d_{\min} = 3$ $R_c$). The final stages of the process 
are similar to what is  observed during pulling simulations, where a set of two NPs must come into contact with the water molecules of the other droplet to initiate coalescence. Unlike the results observed during the pulling simulations, this evolution happens without an increase of the droplets radius of gyration. 

This difference is due to the pulling parameters chosen, which allow the system to sample two different mechanisms during coalescence. 
The parameters chosen for the pulling simulations are such that the system 
remains in the advection regime. This can be highlighted with the use of the P\'eclet number~\cite{Huysmans2005}
\begin{equation}
Pe = \frac{v L_c }{D} \, ,
\end{equation}
where $v$ is the effective advection velocity, $L_c$ a characteristic length, and $D$ the diffusion coefficient.
Considering the pulling velocity in the center of mass reference frame, $v_P$, the DPD characteristic length, $R_c$, 
and the simulated water self-diffusion coefficient $D_{\textrm{sim}}$, 
one obtains $P_e = \frac{v_P R_c}{D_{\textrm{sim}}} = \frac{0.002 R_c/\tau \times R_c }{0.0063 R_c^2/\tau} \equiv 0.3$.
This value for the P\'eclet number indicates that during the pulling simulations the system is dominated 
by the pulling velocity. On the contrary, in the ABMD simulations, the evolution follows the natural fluctuations 
of the system, a regime dominated by diffusion by design.

Our results suggest that when pulling simulations are considered, the coalescence process depends on parameters 
such as NP contact angles and concentrations. On the contrary, the results obtained from ABMD simulations do not show 
such dependencies.
We illustrate this difference with the example in Fig.~\ref{fig4} (left panel) where we keep the interfacial area per NP 
constant (160 NPs) and consider spherical 60JP NPs. In this figure the contact point (C) corresponds 
to the first NP-NP interaction between the two droplets and the merging point (M) identifies
the formation of the water bridge between the two droplets. In Fig.~\ref{fig4} we illustrate that both C and M are different 
in either pulling or ABMD simulations due to the  shape deformation of the system.
The system seeks to keep the NP contact angles as close as possible 
to their equilibrium value (data consistent with Table~\ref{Tab-characteristic}). 
Our simulations suggest that under these conditions, the droplets deform their shape and increase the water-oil 
interfacial area rather than allowing the NPs to change their contact angle or desorb from the interface. 
The increased interfacial area, not covered by NPs as shown in Fig.~\ref{fig4}, favours coalescence.

The differences between the results obtained between ABMD and pulling simulations also increase when the 
surface density of the NPs on the droplets increases. We illustrate this difference with the example 
in Fig.~\ref{fig4} (right panel), where we increase the interfacial area per NP (from 160 NPs to 165 NPs) and consider 
spherical 60JP NPs. Under these conditions, the NP density becomes sufficiently high that the droplet 
cannot create a large NP-free area (as we observed for 160 NPs) to facilitate coalescence. 
When the distance between the two droplets decreases, one NP on the incoming droplet 
comes in contact with the solvent shell and then the water molecules from the other droplet.
We find it interesting that in this scenario coalescence initializes in an area different from 
the initial interaction area (yellow spheres in Fig.~\ref{fig4}). This NP is different 
from the NP that induces the compression of the neighbour droplet (red spheres in Fig.~\ref{fig4}). 
The contact point (C) and the merging point (M) are different and the merging does not require the contact of two NPs 
originally on the two separable droplets.\\

In the ESI, we report the results obtained when homogeneous NPs were simulated.
The results are consistent with pulling simulations reported previously~\cite{FanSM2012}. The key point is 
the difference in the interaction 
between the NP on the incoming droplet, and the other droplet. Due to the homogeneous distribution of polar and apolar beads 
on the NP surface, the homogeneous NP from the incoming droplet can adsorb simultaneously on the two interfaces, yielding the same 
contact angle on both droplets. As a consequence, in both pulling and ABMD simulations, the last step of the coalescence mechanism 
differs from the one observed with spherical Janus NPs. When homogeneous NPs are used, the coalescence requires only the contact of one NP to induce
the merging of the two water droplets.\\

\begin{figure*}
\begin{center}
\includegraphics[width=0.45 \textwidth, angle=-0]{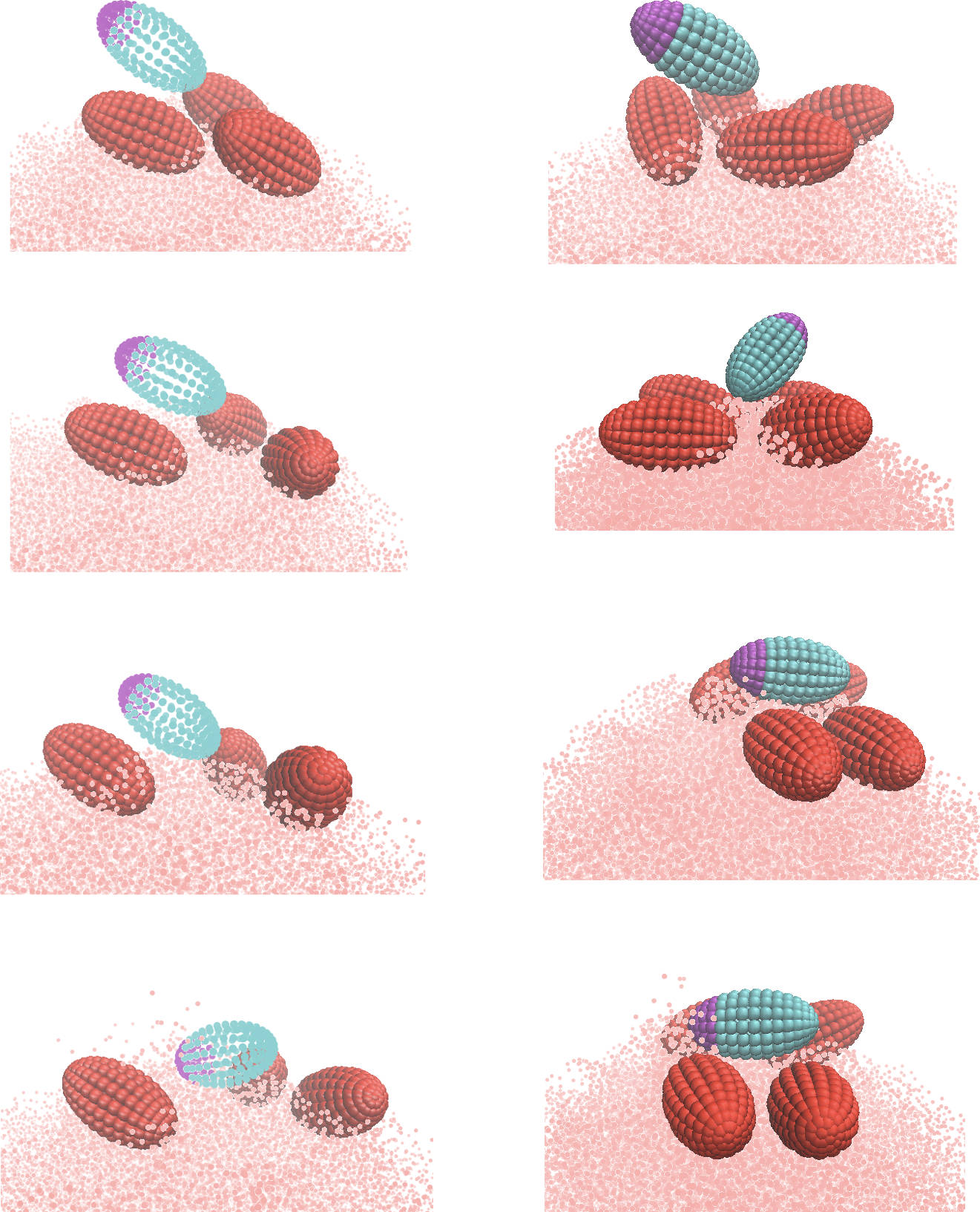}
 \caption{
 Detailed sequence of simulation snapshots representing typical coalescence processes of two 
 water droplets stabilized by 160 ellipsoidal nanoparticles of type 70JP during pulling (left panel) 
 and ABMD simulations (right panel). Green and purple spheres represent polar and apolar beads, respectively. 
 Considering the pulling simulation, one NP on the incoming droplet 
 interacts first with a NP on the other droplet before it starts interacting with the solvent shell around the water molecules. 
 In ABMD simulation, the incoming NP interacts almost immediately with the solvent shell. The rest of the coalescence process is similar 
 to what is observed for both pulling and ABMD simulations. As described in the text, the longitudinal orientation of the 
 Janus ellipsoidal NPs yields a peculiar coalescence mechanism.
}
\label{fig6}
\end{center}
\end{figure*}

\begin{figure*}
\includegraphics[width=1.0 \textwidth, angle=-0]{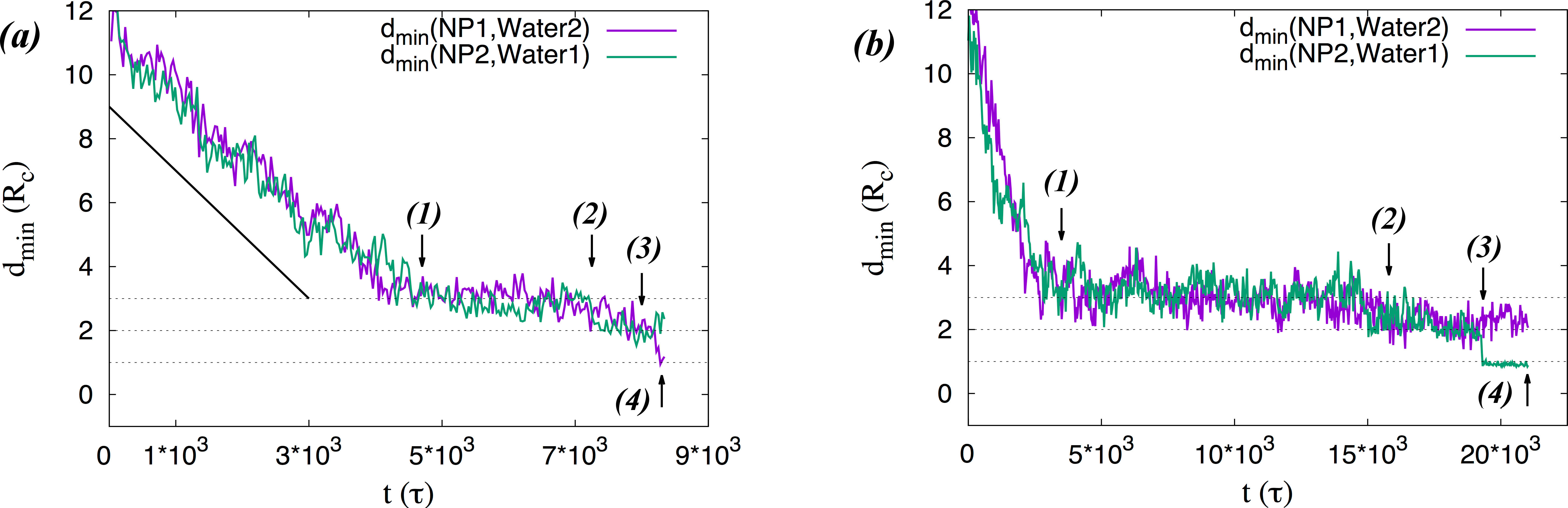}
 \caption{
 Temporal evolution of the minimal distance $d_{\min}$ between the apolar extremity of the incoming droplet (NP1 or NP2) and the water 
 molecules from the other droplet (Water2 or Water1, respectively). Two water droplets stabilized with 160 ellipsoidal nanoparticles of type 70JP 
 in decane solvent are considered. During the pulling simulation (panel a) the distance decreases 
 consistently with the simulation pulling rate (black plain line). The distance 
 $d_{\min}$ reaches a plateau (1) when the incoming NP interacts with a NP belonging to the other droplet. 
 The incoming NP then interacts with the solvent shell surrounding the droplet (2) and the water molecules (3). 
  The last step differs significantly from the one described for spherical NP, 
 as described in the text (4). 
Considering the ABMD simulation (panel b), the distance $d_{\min}$ does not show a clear NP-NP interaction plateau.
 The incoming NP interacts with the solvent shell surrounding the second droplet and the water molecules, 
 \textit{i.e.} $2 R_c \leq d_{\min} \leq 3 R_c$. 
 The last step is similar to the one described for pulling simulations.
 }
\label{fig7}
\end{figure*}

We also considered the impact of different NP shape on the coalescence mechanisms, focusing 
our attention on ellipsoidal NP of aspect ration $c/b = 2$ (242 beads around the ellipsoidal NP). 
In Fig.~\ref{fig5} we first consider the qualitative evolution of the system. As in the case  
with spherical NP, the ABMD simulation does not alter the shape of the droplet, while the pulling simulation induces 
an increase of the droplets gyration radius. This is visually investigated in Fig.~\ref{fig6} 
and more quantitatively shown in Fig.~\ref{fig7}.
The compression of the droplet observed during pulling simulations can be explained in a first instance by the same mechanism we previously described, 
\textit{i.e.}, the pronounced NP-NP interaction due to the non-diffusive behaviour of the system. The incoming NP interacts with the solvent shell around the water molecules and eventually comes into contact with the water molecules themselves.
However, the last step of the coalescence mechanism differs from 
the one observed with spherical Janus NPs (cf. ESI for visual inspection). When the droplets are stabilized by ellipsoidal NPs, 
the coalescence process does not require the contact of two NPs (one on each droplet). 
Due to the longitudinal orientation of the ellipsoidal NP with respect to the interface, 
contact happens between the apolar NP beads (green spheres in Fig.~\ref{fig6}) on the incoming droplet and the 
water molecules of the second droplet. When this contact is secured, water molecules from the second droplet 
can travel a short distance and interact with the polar NP beads (purple spheres in Fig.~\ref{fig6}), originally 
exposed to the decane solvent. Once this interaction is established, a molecular bridge is formed between 
the two droplets that leads to coalescence. It is worth repeating that this coalescence mechanism is facilitated by the 
longitudinal orientation of the Janus ellipsoidal NPs on the droplet interface.

\section{Conclusions}
In this paper, we considered Adiabatic Biased Molecular Dynamics simulations to investigate the early stage 
of the coalescence process between two water droplets in decane solvent at the mesoscopic level. 
The droplets are stabilized with various NPs (Janus and homogeneous) of varying contact angles and shapes, and different NP densities. 
In ABMD simulations, the system is forced to coalesce \textit{adiabatically}. This allows us to study 
how the ability of the NPs to stabilize Pickering emulsions is discriminated by NP mobility and the associated 
caging effect.
We compared how the coalescence process differs when one considers pulling simulations with sufficiently high steering
speeds. When ABMD simulations are considered, we showed that the coalescence mechanism does not depend on parameters 
such as NP contact angles and concentrations. However, it is sensitive to the NP type (Janus NP vs. Homogeneous NP) and 
shape (spherical vs. ellipsoidal). The coalescence occurs when a water bridge is formed between the two 
droplets, which drives the subsequent flow of water molecules. The formation of the water bridge depends on the shape of the NPs used as stabilizers. 
Considering spherical NPs, the water bridge is formed with two NPs (each belonging to different droplets) that interact 
with water molecules from the other droplet. This differs from ellipsoidal NPs of aspect ratio $c/b = 2$. Due to 
the longitudinal orientation of the ellipsoidal NP with respect to the interface, water molecules from the second droplet 
can interact more easily with the polar NP beads on the incoming droplet originally exposed to the decane
solvent. They form a molecular bridge with a single ellipsoidal NP between the two droplets that leads to coalescence.

When pulling simulations are considered, we observed a wider range of coalescence processes. When the system remains 
in the advection regime, as defined by the P\'eclet number, the droplet has the ability to undergo 
large elastic deformations before coalescing. This is due in a first instance to the NP-NP interactions, which are very 
sensitive to the NP density on the droplets. This is related to the caging effect associated to NP mobility. 
The collision speed also influences the alteration of the shape of the water droplet, as the coalescence process 
is sensitive to the time scale associated to the liquid film drainage. These different parameters, associated with the fact 
that the system tries to keep the NP contact angles as close as possible to their equilibrium value when the shape 
of the system changes, are responsible for a wide range of coalescence processes.

These results are consistent with the recent microfluidic collision experiment of Zhou et al.~\cite{Zhou2015} 
studying head-to-head moving water droplets in a microfluidic chip.
Following preliminary investigations on droplet coalescence~\cite{Liao2010,Leal2004}, 
the authors analysed the coalescence process 
using the droplet contact time and the liquid film drainage time (\textit{i.e.}, the time interval from droplet 
contact to fusion). They defined a coalescence percentage to indicate how coalescence is prone to occur under different 
flow conditions. However, as the droplets have the opportunity to detach from each other when droplet velocity is too high, 
the coalescence percentage can eventually reach a zero value. The authors highlighted that droplets coalesce 
effectively at low velocities and suggested that this is related to the importance of the liquid film drainage between 
two approaching droplets and the droplet faculty to undergo elastic deformation.
Our numerical analysis allows us to investigate the early stage of the coalescence process 
at a mesoscopic scale and to propose detailed coalescence mechanisms, the observation of which 
remains at a scale inaccessible through experiments. Although  
the present work only considers frontal droplet-droplet collisions, our results highlight the importance of droplet speed 
in modulating the coalescence process.
\\

To go further, it would be interesting to modify the chemistry of the NPs to study their ability to enhance or 
to perturb emulsion stability. Considering both ABMD and pulling simulations, we showed that the two water droplets start 
coming into contact through  NP-NP interactions. 
Adjusting the apolar-apolar interaction between NPs might have an impact on 
the coalescence process, perhaps increasing or shortening  
the NP-NP interaction time and manipulating the effective interaction with the solvent shell. 
Similarly, it would be interesting to study the effect of particle diameters on the coalescence process. Given a sufficiently 
high NP density around the droplets, increasing the diameter of the Janus NPs would also increase the NP-NP interaction time 
without changing the NP chemistry.
%

\section*{acknowledgement}

We acknowledge X-C. Cuong, V. Garbin, and L. Botto for useful discussions. 
F.S. thanks M. Salvalaglio for fruitful discussion 
concerning the ABMD algorithm. Via our membership of the UK's HEC Materials Chemistry Consortium, 
which is funded by EPSRC (EP/L000202), this work used the ARCHER UK National Supercomputing Service 
(http://www.archer.ac.uk).

\bibliography{rsc} 
\bibliographystyle{rsc} 

\pagebreak
\widetext
\begin{center}
\textbf{\large Numerical analysis of Pickering emulsion stability: insights from ABMD simulations\\
			   Supplemental Information}
\end{center}
\setcounter{equation}{0}
\setcounter{figure}{0}
\setcounter{table}{0}
\setcounter{page}{1}
\makeatletter
\renewcommand{\theequation}{S\arabic{equation}}
\renewcommand{\thefigure}{S\arabic{figure}}
\renewcommand{\bibnumfmt}[1]{[S#1]}
\renewcommand{\citenumfont}[1]{S#1}
\section{Droplet characteristic}

\begin{figure}[b]
\includegraphics[width=0.7 \textwidth, angle=-0]{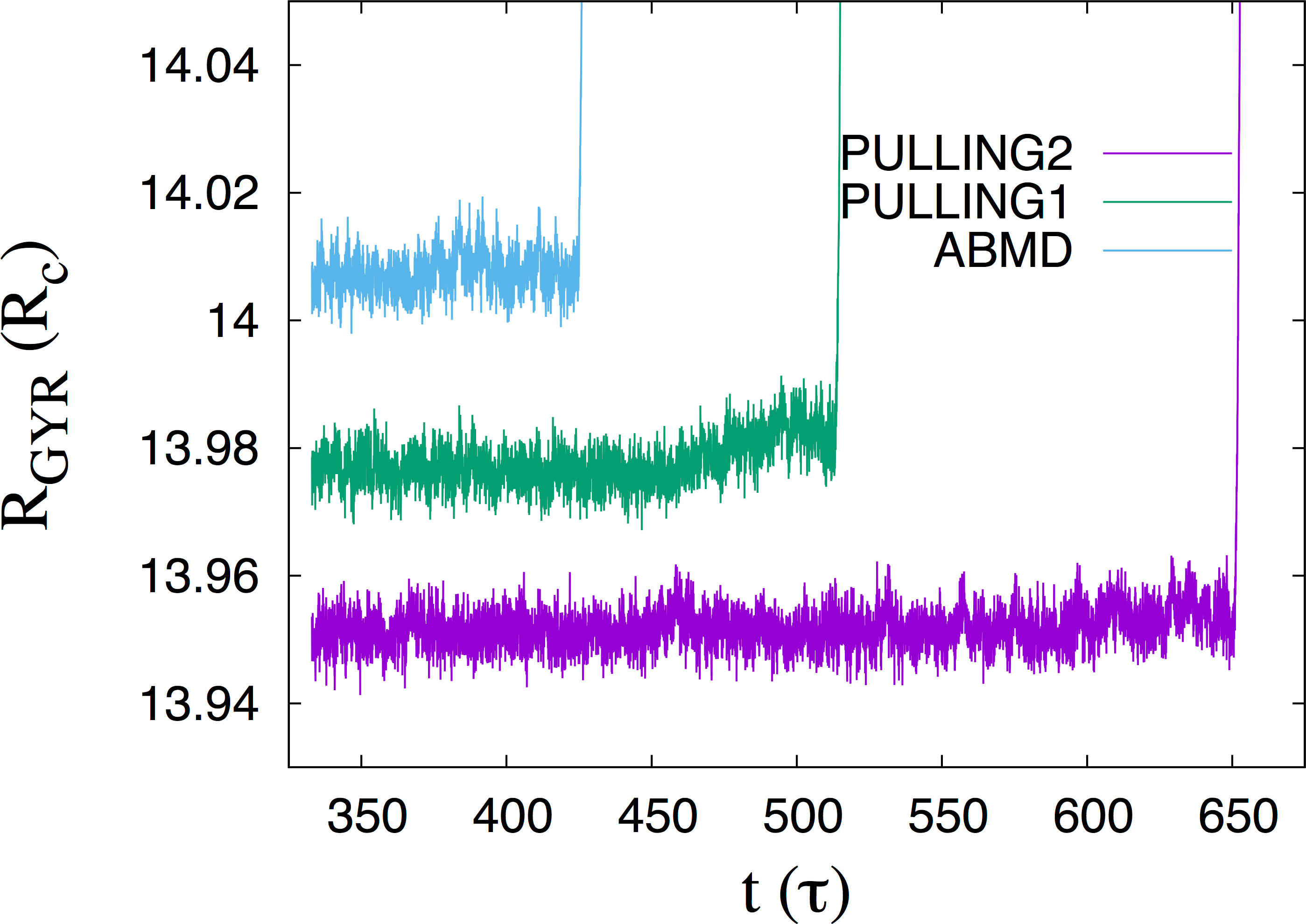}
 \caption{ Temporal evolution of the radius of gyration, $R_{GYR}$, of one incoming droplet during 
  pulling and ABMD simulations. The temporal evolutions of $R_{GYR}$ are shifted arbitrarily along the 
  ordinate axis for clarity. We use the notation: Pulling1  $\big( v_P = 0.001$ $R_C$/time step, $k_P = 5000$ $k_BT/R_C^2 \big)$, 
  Pulling2 $\big( v_P = 0.0005$ $R_C$/time step, $k_P = 5000$ $k_BT/R_C^2 \big)$, 
  and ABMD $\big( \alpha = 1000$ $k_BT/R_C^2$, $S_{target}=39$ $R_C \big)$.}
\label{figS1}
\end{figure}

We consider nanoparticle (NP) types and concentrations  that  are expected to be
(1) strongly effective at preventing droplet coalescence, and (2)  maintain the shape of the droplet spherical. 
As described in prior work~\cite{FanSM2012}, the first criterion can be related to the interfacial area per NP, $\mathcal{A}_{NP}$. 
The NPs are not expected to be strongly effective at preventing droplets coalescence when $\mathcal{A}_{NP}$ is larger 
than $30$ $R_c^2$, independently on whether water or decane droplets are considered. 
Considering the values for $\mathcal{A}_{NP}$ reported in Table 2 in the main text, water droplets stabilized with $144$ NPs 
have an interfacial area per NP $\mathcal{A}_{NP} > 30$ $R_c^2$. These systems do not show high stability, as quantitatively shown in Fig.~\ref{figS1} where 
we follow the temporal evolution of the radius of gyration, $R_{GYR}$, of one water droplet stabilized with 144 spherical NPs 
of type 55JP. The temporal evolutions of $R_{GYR}$ are shifted arbitrarily along the ordinate axis for clarity.
In both pulling and ABMD simulations, $R_{GYR}$ remains constant as the distance between the two droplets decreases.
The departure from the plateau indicates the merging of the two droplets. This merging happens without significant increase of $R_{GYR}$, 
indicating that the resistance to coalescence provided by the NPs is minimal.

\begin{figure}
\includegraphics[width=0.8 \textwidth, angle=-0]{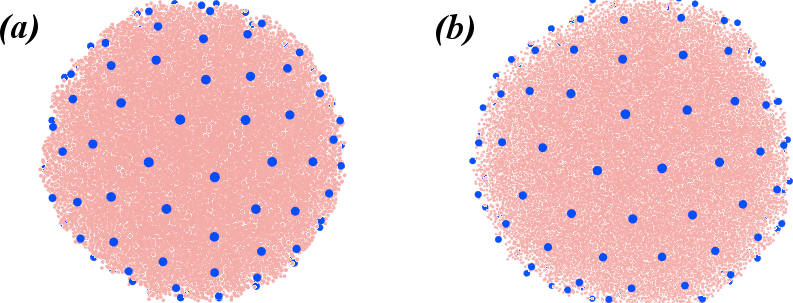}
 \caption{Simulation snapshots representing one water droplet stabilized with either $144$ 
 (panel a) or $170$ spherical NPs (panel b) of type 55JP. We show the positions of the NP centers (blue points), but not he NPs for clarity.}
\label{figS2}
\end{figure}

The NPs can affect the shape of the droplet at small interfacial area per NP, $\mathcal{A}_{NP}$, 
because the system tries to keep the NP contact angles $\theta_C$ as close as possible to their equilibrium 
value (cf. evolution of the contact angle $\theta_C$ in Table 2 
in the main text). In Fig.~\ref{figS2} we compare water droplets 
stabilized with either $144$ spherical NPs (left panel) or $170$ spherical NPs (right panel) of type 55JP. 
Visual inspection demonstrates
the alteration of the droplet shape when $170$ NPs are used. 
In correspondence to the changes described in Fig.~\ref{figS2}, the changes in relative shape anisotropy, reported in 
Table 2 in the main text, are of one order of magnitude of difference.
%

\section{Water droplets stabilized by homogeneous NPs}

\begin{figure}[h]
\includegraphics[width=0.8 \textwidth, angle=-0]{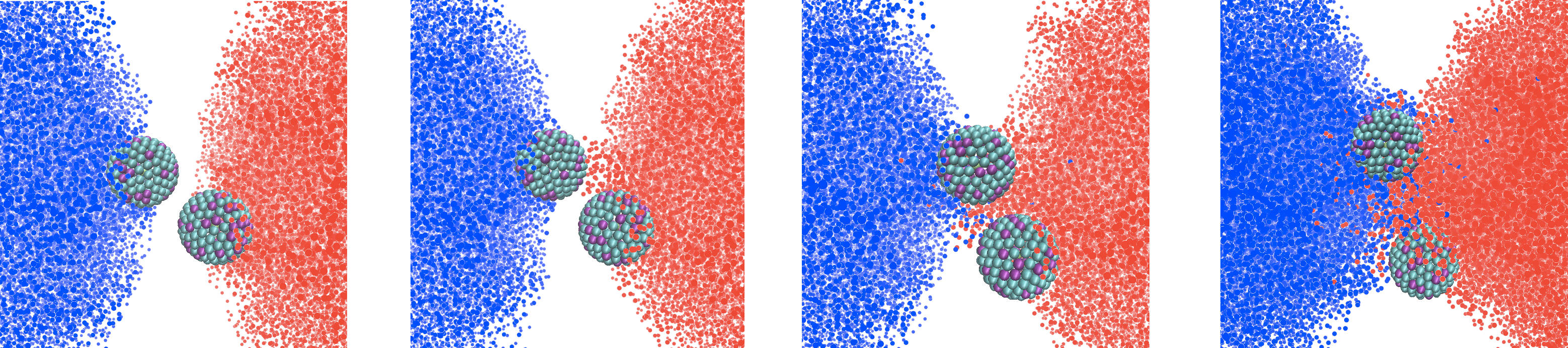}
 \caption{ Detailed sequence of simulation snapshots representing typical coalescence process of two water droplets (red or blue) in decane solvent. The droplets are stabilized with 160 spherical NPs of type 75HP. The results are obtained using ABMD simulations. Green and purple spheres represent polar and apolar beads, respectively.
 The distance between the two droplets decreases from left to right. 
 The coalescence process only requires  the contact of one NP to induce the merging of the two water droplets.}
\label{figS3}
\end{figure}

We  considered the ability of homogeneous (HP) NPs to stabilize Pickering emulsions.
As we explained in the main text, due to the homogeneous distribution of polar and apolar beads 
on the NP surface, one homogeneous NP can adsorb simultaneously on  two interfaces yielding the same 
contact angle on both droplets. The results obtained during the present coalescence simulations are consistent with pulling simulations reported previously~\cite{FanSM2012}.
In both pulling and ABMD simulations the last step of the coalescence mechanism 
differs from the one observed with spherical Janus NPs. In Fig.~\ref{figS3} we show detailed sequence of simulation 
snapshots representing typical collision processes of two water droplets in decane solvent. The droplets are stabilized by 160 spherical NPs 
of type 75HP (contact angle $\theta_c = 101.7 \pm 3.5$). The results are obtained with  ABMD simulations. The key point is the interaction between the incoming NP and the second droplet. Indeed, the coalescence process requires only the contact 
of one NP to induce the merging of the two water droplets.

\section{Coalescence mechanism for ellipsoidal NPs}
\begin{figure}
\includegraphics[width=0.3 \textwidth, angle=-0]{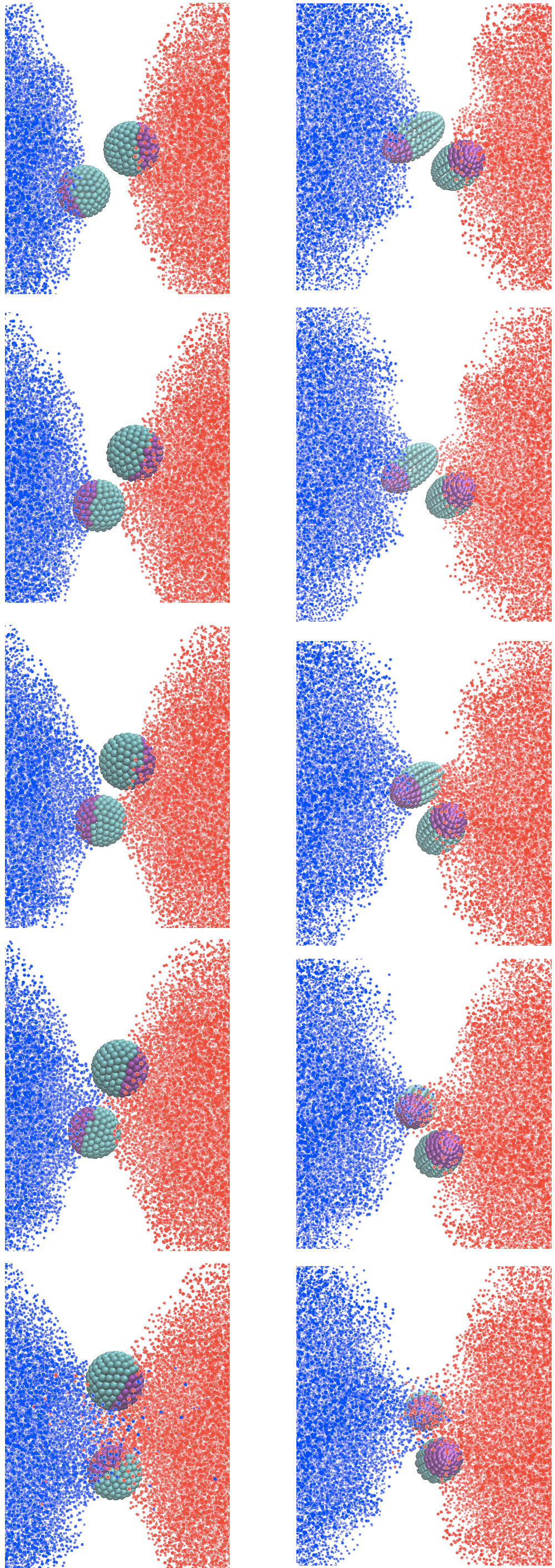}
 \caption{ Detailed sequences of simulation snapshots representing typical coalescence process of two water droplets (red or blue) in decane 
 solvent. The droplets are stabilized with either 160 spherical NPs of type 60JP (left panel) 
 or 160 ellipsoidal NPs of type 70JP (right panel). 
 The results are obtained using ABMD simulations. Green and purple spheres represent polar and apolar beads, respectively.
 The distance between the two droplets decreases from top to bottom. When the droplets are stabilized 
 by ellipsoidal NPs (right panel), the coalescence process only requires  the contact of one NP to bridge the two interfaces ans to induce the merging of the two water droplets.}
\label{figS4}
\end{figure}

We show in Fig.~\ref{figS4}  detailed sequences of simulation snapshots representing typical coalescence processes 
of two water droplets stabilized with either 160 spherical NPs of type 60JP (left panel) or 160 ellipsoidal NPs of type 70JP (right panel). 
The results are obtained using ABMD simulations. When the droplets are stabilized by ellipsoidal NPs, the coalescence process does not require 
the contact of two NPs from the two merging droplets. This is due to the longitudinal orientation of the ellipsoidal NP with respect 
to the interface, as explained in the main text.
%



\end{document}